\documentclass[journal]{IEEEtran}
\usepackage{amssymb}
\usepackage[cmex10]{amsmath}
\usepackage{stfloats}
\usepackage{graphicx}
\usepackage{subfigure}
\usepackage{tabularx}
\usepackage{epsfig,epsf,color,balance,cite}
\usepackage{verbatim}
\usepackage{url}
\usepackage{bm}
\usepackage{epstopdf}
\usepackage{algorithm}
\usepackage{algorithmic}
\usepackage{cuted}

\usepackage{algorithm}
\usepackage{algorithmic}

\usepackage{color}
\definecolor{myc1}{rgb}{0,0,0}

\definecolor{myc2}{rgb}{0,0,0}
\begin{document}

\title{MSE-Based Transceiver Designs for RIS-Aided Communications with Hardware Impairments}

\author{
\IEEEauthorblockN{Jingwen Zhao, Ming Chen,  Cunhua Pan, Zhiyang Li, Gui Zhou, Xiao Chen
                  \vspace{-3em}}
\thanks{This work was supported in part by the National Key Research and Development Program under grant 2018YFB1801905, in part by the National Natural Science Foundation of China (NSFC) under grant 61960206005 \& 61960206006, in part by the NSFC under Grant 62101273.(\itshape Corresponding author: Jingwen Zhao.)}
\thanks{J. Zhao, M. Chen, Z. Li and C. Pan are with the National Mobile Communications Research Laboratory, Southeast University, Nanjing 210096, China, email: \{zhaojingwen, chenming, lizhiyang, cpan\}@seu.edu.cn.}
\thanks{G. Zhou is with School of Electronic Engineering and Computer Science, Queen Mary University of London, E1 4NS, U.K., email: g.zhou@qmul.ac.uk.}
\thanks{X. Chen is with the College of Artificial Intelligence, Nanjing University of Information Science and Technology, Nanjing, 210044, China, email: X.Chen@nuist.edu.cn.}
}

\maketitle 

\begin{abstract}
It is challenging to precisely configure the phase shifts of the  reflecting elements at the reconfigurable intelligent surface (RIS) due to inherent hardware impairments
(HIs). 
In this paper, 
the mean square error (MSE) performance is investigated in an RIS-aided single-user multiple-input multiple-output (MIMO) communication system  with transceiver HIs and RIS phase noise.
We aim to jointly optimize  the transmit precoder, linear received equalizer, and RIS reflecting matrices to minimize the MSE.
To tackle this problem,  an iterative algorithm  is proposed, wherein the beamforming matrices are alternately optimized.
Specifically, for the beamforming  optimization subproblem, we derive the closed-form expression of the optimal precoder and equalizer matrices.
Then, for the phase shift optimization subproblem,  an efficient algorithm based on the majorization-minimization (MM) method is proposed. 
Simulation results show that the proposed MSE-based RIS-aided transceiver scheme dramatically outperforms the conventional system algorithms that do not consider HIs at both the transceiver and the RIS.

\end{abstract}

\IEEEpeerreviewmaketitle

\section{Introduction}
Thanks to its appealing advantages of enhancing the system energy efficiency, reconfigurable intelligent surface (RIS) has attracted extensive research attention in both academia and industry \cite{9}.
Specifically, by deploying a large number of passive reflecting elements, RIS can  adjust the phase shifts of the reflecting elements  to enhance the transmission signal power and cancel interference.
Different from conventional amplify-and-forward  relays, RIS  just passively reflects the incident signals instead of receiving or transmitting the signals, which greatly reduces the energy consumption \cite{10}. 
For  RIS-aided communication systems,
the  transceiver design has been studied in  \cite{2,944,966,3}. 
Specifically,
the
authors in \cite{944} investigated the weighted sum secrecy rate maximization problem by
jointly optimizing the active transmit beamforming and passive reflecting
phase shifts.   
Furthermore, 
the authors in \cite{966} and \cite{3} considered similar design problems for RIS-aided single-user multiple-input single-output (MISO),
and single-user multiple-input multiple-output (MIMO) systems, respectively, with an aim of minimizing the   mean square error (MSE).

However,  all the above contributions assumed  ideal transceiver hardware and perfect RIS phase-shifting  without considering hardware impairments (HIs).
The transceiver HIs and RIS phase errors always exist in actual systems, limiting channel capacity, particularly in the high-power region, and degrading beamforming gains.
Unfortunately, the compensation algorithms cannot completely eliminate these impairments because of the time-varying hardware characteristics.
As a result,
several researches have studied the impact of HIs at both the transceiver and the RIS in various RIS-aided scenarios.
The authors in \cite{5} investigated the joint calibration of the
direction-independent and direction-dependent phase errors
for an RIS-aided millimeter-wave system
 on the system performance.
Besides, 
an RIS-aided MISO system
was studied based on statistical channel state information (CSI)  in \cite{6} to maximize the system sum rate
with the consideration of HIs at both transceiver and the RIS.
However, in these  contributions \cite{5,6,900}, users are equipped with single antenna and  the impact of
transceiver HIs and RIS phase noise on an RIS-aided MIMO system has not been investigated. 
Therefore, this work focuses on the downlink transmission of an RIS-aided single-user MIMO system, wherein there exists transceiver HIs and RIS phase noise.
It is challenging to jointly design the precoder and equalizer matrices at the MIMO transceiver and phase shifts of the RIS, which has not been studied in the existing works, to the best of our knowledge. 

In this paper,  we study the transceiver design for  an RIS-aided single-user  MIMO communication system with  transceiver HIs and RIS phase noise.
We jointly optimize the transmit precoder, the received equalizer, and the RIS reflecting matrices to minimize the MSE of this system.
The proposed optimization problem is tackled by using alternating optimization (AO) method based on  Lagrange dual and majorization-minimization (MM) techniques. 
Simulation results show  that the performance advantage of the proposed transceiver design scheme and reveal the importance of considering HIs impact on the transceiver design.

\section{System Model and Problem Formulation}
We consider an RIS-aided downlink communication system consisting of one $N_\mathrm{t}$-antenna base station (BS), one $M$-element RIS, and one $N_\mathrm{r}$-antenna  user.
Let $\mathbf{s}\in\mathbb{C}^{d\times 1}$ denote the data streams from the BS satisfying $\mathbb{E}[\mathbf{s}]=\mathbf{0}, \mathbb{E}[\mathbf{s}\mathbf{s}^\mathrm{H}]=\mathbf{I}_d$, where $d$ is the number of data streams.
The undistorted transmit vector $\mathbf{x}\in\mathbb{C}^{N_\mathrm{t}\times 1}$ is given by
\begin{equation}\label{a88}
	\mathbf{x}=\mathbf{W}\mathbf{s},
\end{equation}
where $\mathbf{W}\in\mathbb{C}^{N_\mathrm{t}\times d}$ is the  precoder matrix at the BS.
The signal transmitted by the BS is given by
\begin{equation}\label{a1}
\tilde{\mathbf{x}}=\mathbf{x}+\mathbf{z}_\mathrm{s},
\end{equation}
where  $\mathbf{z}_\mathrm{s}\in\mathbb{C}^{N_\mathrm{t}\times 1}$ represents the transmit distortion noise which is independent of $\mathbf{s}$. 
More specifically, $\mathbf{z}_\mathrm{s}$ follows the independent zero-mean Gaussian random distribution, i.e., $\mathbf{z}_\mathrm{s}\sim\mathcal{CN}(\mathbf{0},\kappa_\mathrm{s}\text{diag}\{\mathbf{W}\mathbf{W}^\mathrm{H}\})$, where $\kappa_\mathrm{s}\in(0,1)$ denotes the normalized variance of the transmit distortion noise.
We adopt the channel  estimation methods for RIS-aided systems in  \cite{913} and thus  we assume that the perfect CSI is available at the transmitter.
The received signal at the   user is expressed as
\begin{equation}\label{a2}
\begin{aligned}
\mathbf{y}&=(\mathbf{H}_\mathrm{d}^\mathrm{H}+\mathbf{H}_\mathrm{r}^\mathrm{H}\mathbf{\Theta}\mathbf{\hat{\Theta}}\mathbf{H}_\mathrm{t})\tilde{\mathbf{x}}+\mathbf{z}_\mathrm{d}+\mathbf{n}_\mathrm{d}
\triangleq \tilde{\mathbf{y}}+\mathbf{z}_\mathrm{d}+\mathbf{n}_\mathrm{d},
\end{aligned}
\end{equation}
where $\mathbf{H}_\mathrm{d}\in\mathbb{C}^{N_\mathrm{t}\times N_\mathrm{r}}$ is the user-to-BS channel,  $\mathbf{H}_\mathrm{r}\in\mathbb{C}^{M\times N_\mathrm{r}}$ is the user-to-RIS channel, and $\mathbf{H}_\mathrm{t}\in\mathbb{C}^{M\times N_\mathrm{t}}$ is the BS-to-RIS channel.  $\mathbf{\Theta}=\text{diag}\{e^{j\theta_1},e^{j\theta_2},\cdots,e^{j\theta_M}\}$ is the diagonal reflection matrix of the RIS with $\theta_m\in [0,2\pi]  $~$ (m=1,2,\cdots,M) $,  
and $ \mathbf{n}_\mathrm{d} \sim\mathcal{CN}(0,\sigma_\mathrm{n}^2\mathbf{I})$ represents the thermal additive white Gaussian noise (AWGN). $\mathbf{z}_\mathrm{d}\in\mathbb{C}^{N_\mathrm{r}\times 1}$ denotes the received distortion noise independent of $\tilde{\mathbf{y}}$
and the variance of $\mathbf{z}_\mathrm{d}$ is proportional to the undistorted received signal power, i.e., $\mathbf{z}_\mathrm{d}\sim\mathcal{CN}(\mathbf{0},\kappa_\mathrm{d}\text{diag}\{\mathbb{E}\left[\tilde{\mathbf{y}}\tilde{\mathbf{y}}^\mathrm{H}\right]\}$, where $\kappa_\mathrm{d}\in(0,1)$ is the  normalized ratio of distorted noise power to undistorted received signal power.
$ \mathbf{\hat{\Theta}}=\text{diag}\{e^{j\varepsilon_1},e^{j\varepsilon_2},\cdots,e^{j\varepsilon_M}\}$  is the random phase noise matrix , wherein $\varepsilon_m$ is each RIS element's phase noise caused by RIS HIs.


It is assumed that the linear equalizer matrix $\mathbf{C}\in\mathbb{C}^{d\times N_\mathrm{r}}$ is used to equalize the received signal. The estimated signal at the user is given by
	\vspace{-0.55em}
\begin{align}\label{a3}
\!\!\!\hat{\mathbf{s}}\!\!=\!\!\mathbf{C}\mathbf{y}
=\!\!\mathbf{C}[(\mathbf{H}_\mathrm{d}^\mathrm{H}+\mathbf{H}_\mathrm{r}^\mathrm{H}\mathbf{\Theta}\mathbf{\hat{\Theta}}\mathbf{H}_\mathrm{t})\left(\mathbf{W}\mathbf{s}+\mathbf{z}_\mathrm{s}\right)\!+\!\mathbf{z}_\mathrm{d}+\mathbf{n}_\mathrm{d}].	\vspace{-1.85em}
\end{align}
Thus, the MSE of this system is defined as follows
\begin{align}\label{ww}
	& \text{MSE}=\mathbb{E}\{\Vert\hat{\mathbf{s}}-\mathbf{s}\Vert_{2}^{2}\}=\mathbb{E}_{\mathbf{\hat{\Theta}},\mathbf{s}}\{\text{tr}[(\hat{\mathbf{s}}-\mathbf{s})(\hat{\mathbf{s}}-\mathbf{s})^{\mathrm{H}}]\}.
\end{align}

We assume that the phase noise variable $\varepsilon$  follows a zero-mean Von Mises Distributions with a concentration parameter $\varsigma$ \cite{8}.
We can  obtain
\begin{align}\label{qqq}
	\text{E}\{e^{j\varepsilon}\}=\frac{I_1(\varsigma)}{I_0(\varsigma)}\triangleq\rho,
\end{align}
where $I_n(\varsigma)$ represents the modified Bessel function of the first kind and order $n$.
\begin{figure*}[!b]
	\vspace{-1.5em}
	\rule[1.5ex]{2.05\columnwidth}{0.5pt}
	\begin{align}\label{a4}
		\text{MSE} & =\textrm{tr}\left\{ \mathbf{C}\left(\mathbf{H}_{\mathrm{d}}^{\mathrm{H}}+\rho\mathbf{H}_{\mathrm{r}}^{\mathrm{H}}\mathbf{\Theta}\mathbf{H}_{\mathrm{t}}\right)\mathbf{W}\mathbf{W}^{\mathrm{H}}\left(\mathbf{H}_{\mathrm{d}}^{\mathrm{H}}+\rho\mathbf{H}_{\mathrm{r}}^{\mathrm{H}}\mathbf{\Theta}\mathbf{H}_{\mathrm{t}}\right)^{\mathrm{H}}\mathbf{C}^{\mathrm{H}}\right\} +\left(1-\rho^{2}\right)\textrm{tr}\left\{ \mathbf{C}\mathbf{H}_{\mathrm{r}}^{\mathrm{H}}\mathbf{\Theta}\textrm{diag}\left\{  \mathbf{H}_{\mathrm{t}}\mathbf{W}\mathbf{W}^{\mathrm{H}}\mathbf{H}_{\mathrm{t}}^{\mathrm{H}}\right\}  \mathbf{\Theta}^{\mathrm{H}}\mathbf{H}_{\mathrm{r}}\mathbf{C}^{\mathrm{H}}\right\} \nonumber \\
		& +\kappa_{\mathrm{s}}\textrm{tr}\left\{ \!\mathbf{C}\left(\mathbf{H}_{\mathrm{d}}^{\mathrm{H}}+\rho\mathbf{H}_{\mathrm{r}}^{\mathrm{H}}\mathbf{\Theta}\mathbf{H}_{\mathrm{t}}\right)\textrm{diag}\left\{  \mathbf{W}\mathbf{W}^{\mathrm{H}}\right\} \left(\mathbf{H}_{\mathrm{d}}^{\mathrm{H}}+\rho\mathbf{H}_{\mathrm{r}}^{\mathrm{H}}\mathbf{\Theta}\mathbf{H}_{\mathrm{t}}\right)^{\mathrm{H}}\!\mathbf{C}^{\mathrm{H}}\right\} \!+\!\kappa_{\mathrm{s}}\left(1-\rho^{2}\right)\!\textrm{tr}\left\{ \!\mathbf{C}\mathbf{H}_{\mathrm{r}}^{\mathrm{H}}\mathbf{\Theta}\textrm{diag}\left\{ \mathbf{H}_{\mathrm{t}}\textrm{diag}\left\{ \mathbf{W}\mathbf{W}^{\mathrm{H}}\right\} \right.\right.\nonumber \\
		& \left.\left.\mathbf{H}_{\mathrm{t}}^{\mathrm{H}}\right\} \mathbf{\Theta}^{\mathrm{H}}\mathbf{H}_{\mathrm{r}}\mathbf{C}^{\mathrm{H}}\right\} +\kappa_{\mathrm{d}}\textrm{tr}\left\{ \!\mathbf{C}\textrm{diag}\left\{ \left(\mathbf{H}_{\mathrm{d}}^{\mathrm{H}}+\rho\mathbf{H}_{\mathrm{r}}^{\mathrm{H}}\mathbf{\Theta}\mathbf{H}_{\mathrm{t}}\right)\mathbf{W}\mathbf{W}^{\mathrm{H}}\left(\mathbf{H}_{\mathrm{d}}^{\mathrm{H}}+\rho\mathbf{H}_{\mathrm{r}}^{\mathrm{H}}\mathbf{\Theta}\mathbf{H}_{\mathrm{t}}\right)^{\mathrm{H}}\right\} \!\mathbf{C}^{\mathrm{H}}\right\} \!-\textrm{tr}\left\{ \mathbf{C}\left(\mathbf{H}_{\mathrm{d}}^{\mathrm{H}}+\rho\mathbf{H}_{\mathrm{r}}^{\mathrm{H}}\mathbf{\Theta}\mathbf{H}_{\mathrm{t}}\right)\mathbf{W}\right\} \nonumber \\
		& \!\!\!+\!\!\kappa_{\mathrm{d}}\left(\!1\!-\!\rho^{2}\!\right)\!\textrm{tr}\left\{ \!\mathbf{C}\textrm{diag}\left\{ \mathbf{H}_{\mathrm{r}}^{\mathrm{H}}\mathbf{\Theta}\textrm{diag}\left\{ \!\mathbf{H}_{\mathrm{t}} \mathbf{W}\mathbf{W}^{\mathrm{H}} \mathbf{H}_{\mathrm{t}}^{\mathrm{H}}\right\} \!\mathbf{\Theta}^{\mathrm{H}}\mathbf{H}_{\mathrm{r}}\right\} \!\mathbf{C}^{\mathrm{H}}\right\} \!\!-\!\!\textrm{tr}\left\{ \!\mathbf{W}^{\mathrm{H}}\!\left(\mathbf{H}_{\mathrm{d}}^{\mathrm{H}}\!\!+\!\!\rho\mathbf{H}_{\mathrm{r}}^{\mathrm{H}}\mathbf{\Theta}\mathbf{H}_{\mathrm{t}}\right)^{\mathrm{H}}\!\mathbf{C}^{\mathrm{H}}\right\} \!+\!\textrm{tr}\left\{ \sigma_{\mathrm{n}}^{2}\mathbf{C}\mathbf{C}^{\mathrm{H}}\!\!+\!\mathbf{I}+\mathbf{Y}\right\} ,
	\end{align}
\end{figure*}
Then we  derive the mean of the random phase noise matrix as follows
\begin{align}\label{qq}
	\mathbb{E}\{\mathbf{\hat{\Theta}}\}&=\mathbb{E}\{\text{diag}\{e^{j\varepsilon_1},e^{j\varepsilon_2},\cdots,e^{j\varepsilon_M}\}\\
	&=\text{diag}\{\mathbb{E}\{e^{j\varepsilon_1}\}, \mathbb{E}\{e^{j\varepsilon_2}\},\cdots, \mathbb{E}\{e^{j\varepsilon_M}\}\}\nonumber=\rho \mathbf{I}_M.
\end{align}
To simplify the Eq.(5), we introduce an arbitrary square matrix $ \pmb{\varPi}\in \mathbb{C}^{M\times M} $  to represent the combination of  variables $ \mathbf{W},\mathbf{C}$ and $\mathbf{\Theta} $, which can be defined as follows
\small{\begin{equation}\label{yy}
	\pmb{\varPi}=\left[\begin{array}{cccc}
		\pi_{11} & \pi_{12} & \cdots & \pi_{1M}\\
		\pi_{21} & \pi_{22} & \cdots & \pi_{2M}\\
		\vdots & \vdots & \ddots & \vdots\\
		\pi_{M1} & \pi_{M2} & \cdots & \pi_{MM}
	\end{array}\right].
\end{equation}}
From \eqref{qqq} and \eqref{yy},  we can obtain the equivalent phase noise autocorrelation matrix as follows
\small{
	\begin{align}
		\setlength{\arraycolsep}{1.2pt}
		\begin{array}{lc}
			\mathbb{E}_{\mathbf{\hat{\Theta}}}[\mathbf{\hat{\Theta}}\pmb{\varPi}\mathbf{\hat{\Theta}}^{\textrm{H}}]\\=
			\begin{bmatrix}
				\pi_{11}  &   \pi_{12}\mathbb{E}_{\varepsilon}\left[e^{j\varepsilon_{1}-j\varepsilon_{2}}\right]   &   \cdots   &   \pi_{1M}\mathbb{E}_{\varepsilon}\left[e^{j\varepsilon_{1}-j\varepsilon_{M}}\right]   \\
				\pi_{21}\mathbb{E}_{\varepsilon}\left[e^{j\varepsilon_{2}-j\varepsilon_{1}}\right]   &   \pi_{22} &   \cdots    &  \pi_{2M}\mathbb{E}_{\varepsilon}\left[e^{j\varepsilon_{2}-j\varepsilon_{M}}\right]\\
				\vdots  &   \vdots   &   \ddots   &   \vdots  \\
				\pi_{M1}\mathbb{E}_{\varepsilon}\left[e^{j\varepsilon_{M}-j\varepsilon_{1}}\right]   &   \pi_{M2}\mathbb{E}_{\varepsilon}\left[e^{j\varepsilon_{M}-j\varepsilon_{2}}\right]   &  \cdots   &   \pi_{MM} 
			\end{bmatrix}\\
			=\rho^{2}\pmb{\varPi}+\left(1-\rho^{2}\right)\textrm{diag}\{\pmb{\varPi}\}.
		\end{array}
		\label{tt}
\end{align}}

By using \eqref{qqq}, \eqref{qq} and \eqref{tt},
Eq.\eqref{ww}
can be further simplified as \eqref{a4} shown at the bottom of this page, where 
\begin{align*}
	\!\!\mathbf{Y}\!\!=  \!\kappa_{\mathrm{s}}\kappa_{\mathrm{d}}\mathbf{C}\text{diag}\{(\mathbf{H}_{\mathrm{d}}^{\mathrm{H}}\!\!+\!\!\rho\mathbf{H}_{\mathrm{r}}^{\mathrm{H}}\mathbf{\Theta}\mathbf{H}_{\mathrm{t}})\text{diag}\{\mathbf{W}\mathbf{W}^{\mathrm{H}}\!\} (\mathbf{H}_{\mathrm{d}}^{\mathrm{H}}\!\!+\!\!\rho\mathbf{H}_{\mathrm{r}}^{\mathrm{H}}\mathbf{\Theta}\mathbf{H}_{\mathrm{t}})^{\mathrm{H}}\}\mathbf{C}^{\mathrm{H}}
\end{align*}
is the covariance of the transmitted distortion noise and ignored since the multiplication of $\kappa_\mathrm{s}$ and $\kappa_\mathrm{d}$ is very small. 

In this paper, we aim to minimize the MSE in Eq. \eqref{a4} by jointly optimizing the  transmit precoder $\mathbf{W}$, the linear received equalizer $\mathbf{C}$, and the RIS reflecting phase $\mathbf{\Theta}$ while guaranteeing the power constraint at the BS and the unit-modulus constraints at the RIS.
The optimization problem is formulated as follows
\begin{subequations}\label{a5}
\begin{align}	
	&\underset{\mathbf{\Theta},\mathbf{W},\mathbf{C}}{\min}  ~ \text{MSE}\\
	 &\text{s.t.}\ \text{tr}\{\mathbb{E}[\tilde{\mathbf{x}}\tilde{\mathbf{x}}^\mathrm{H}]\}\leq \tau, \label{999}\\ 
	&~~~~\ 0\leq\theta_m\leq2\pi,m=1,\cdots,M.
\end{align}
\end{subequations}
Constraint \eqref{999} is the average transmit power constraint, where $\tau$ is the maximum transmit power at the BS.
It is obvious that Problem \eqref{a5} is  nonconvex, which is difficult to find the optimal solution.
In the following,  we adopt an AO method to  solve this problem.
	\vspace{-0.75em}
\section{Algorithm Design}
To solve Problem \eqref{a5}, we  decompose the original problem into three subproblems.
Specifically,
we firstly update the linear received equalizer $\mathbf{C}$ given transmit precoder $\mathbf{W}$ and RIS reflecting phase shifts $\mathbf{\Theta}$. Then we optimize precoder $\mathbf{W}$  given $\mathbf{C}$ and $\mathbf{\Theta}$.
Finally, the RIS reflecting phase shifts $\mathbf{\Theta}$ is updated given $\mathbf{W}$ and $\mathbf{C}$.
Repeat the above  process until convergence.

\subsection{Transceiver Optimization}
Firstly, we update the linear received equalizer $\mathbf{C}$ with given  precoder $\mathbf{W}$ and  RIS reflecting matrix $\mathbf{\Theta}$.
For convenience, we define 
$\tilde{\mathbf{H}}\triangleq\mathbf{H}_\mathrm{d}^\mathrm{H}+\rho\mathbf{H}_\mathrm{r}^\mathrm{H}\mathbf{\Theta}\mathbf{H}_\mathrm{t}$, $\mathbf{N}_\mathrm{x}\triangleq\tilde{\mathbf{H}}\mathbf{W}\mathbf{W}^\mathrm{H}\tilde{\mathbf{H}}^\mathrm{H}$, $\mathbf{N}_\mathrm{v}\triangleq\tilde{\mathbf{H}}\text{diag}\{\mathbf{W}\mathbf{W}^\mathrm{H}\}\tilde{\mathbf{H}}^\mathrm{H}\!\!\!$, $ \mathbf{N}_{\mathrm{t}}  \triangleq\mathbf{H}_{\mathrm{t}}\mathbf{W}\mathbf{W}^{\mathrm{H}}\mathbf{H}_{\mathrm{t}}^{\mathrm{H}} $, and $ \mathbf{N}_{\mathrm{r}}  \triangleq\mathbf{H}_{\mathrm{t}}\textrm{diag}\left\{ \mathbf{W}\mathbf{W}^{\mathrm{H}}\right\} \mathbf{H}_{\mathrm{t}}^{\mathrm{H}} $.
With given $\mathbf{W}$ and  $\mathbf{\Theta}$, Problem \eqref{a5} can be rewritten as
	\vspace{-0.85em}
\begin{align}	\label{a6}
	\underset{\mathbf{C}}{\min} ~~ \text{MSE}_{c},
\end{align}
where  $	\text{MSE}_{c}$  is given in \eqref{ee}   at the top of the next page.
\begin{figure*}[!t]
	\begin{align}\label{ee}
		\text{MSE}_{c}
		= & \text{tr}\{\!\mathbf{C}\mathbf{N}_{\mathrm{x}}\mathbf{C}^{\mathrm{H}}+\kappa_{\mathrm{s}}\mathbf{C}\mathbf{N}_{\mathrm{v}}\mathbf{C}^{\mathrm{H}}\!+\!\kappa_{\mathrm{d}}\mathbf{C}\text{diag}\{\mathbf{N}_{\mathrm{x}}\}\mathbf{C}^{\mathrm{H}}\!+\!\sigma_{\mathrm{n}}^{2}\mathbf{C}\mathbf{C}^{\mathrm{H}}-\mathbf{C}\tilde{\mathbf{H}}\mathbf{W}-\mathbf{W}^{\mathrm{H}}\tilde{\mathbf{H}}^{\mathrm{H}}\mathbf{C}^{\mathrm{H}}+\mathbf{I}\}\!+\!\left(1-\rho^{2}\right)\textrm{tr}\{\mathbf{C}\mathbf{H}_{\mathrm{r}}^{\mathrm{H}}\mathbf{\Theta}\textrm{diag}\left\{ \mathbf{N}_{\mathrm{t}}\right\} \nonumber\\
		& \mathbf{\Theta}^{\mathrm{H}}\mathbf{H}_{\mathrm{r}}\mathbf{C}^{\mathrm{H}}\!+\!\kappa_{\mathrm{s}}\mathbf{C}\mathbf{H}_{\mathrm{r}}^{\mathrm{H}}\mathbf{\Theta}\textrm{diag}\left\{ \mathbf{N}_{\mathrm{r}}\right\} \mathbf{\Theta}^{\mathrm{H}}\mathbf{H}_{\mathrm{r}}\mathbf{C}^{\mathrm{H}}\!+\!\kappa_{\mathrm{d}}\mathbf{C}\textrm{diag}\{\mathbf{H}_{\mathrm{r}}^{\mathrm{H}}\mathbf{\Theta}\textrm{diag}\left\{ \mathbf{N}_{\mathrm{t}}\right\} \mathbf{\Theta}^{\mathrm{H}}\mathbf{H}_{\mathrm{r}}\}\mathbf{C}^{\mathrm{H}}\}.
	\end{align}
	\vspace{-2.7em}
\end{figure*}
Note that Problem \eqref{a6} is unconstrained. 
Therefore, 
 the optimal $\mathbf{C}^{\mathrm{opt}}$ can be obtained by setting the first-order derivative of $ \text{MSE}_{c} $ with
respect to $ \mathbf{C} $ to zero, as follows
\begin{align}\label{a7}
	\mathbf{C}^{\mathrm{opt}}= & \mathbf{W}^{\mathrm{H}}\tilde{\mathbf{H}}^{\mathrm{H}}[\mathbf{N}_{\mathrm{x}}+\kappa_{\mathrm{s}}\mathbf{N}_{\mathrm{v}}\!+\!\kappa_{\mathrm{d}}\text{diag}\{\mathbf{N}_{\mathrm{x}}\}+\sigma_{\mathrm{n}}^{2}\mathbf{I}+\left(1-\rho^{2}\right)\nonumber\\
	& \!\!\!\!\!\!\mathbf{H}_{\mathrm{\mathrm{r}}}^{\mathrm{H}}\mathbf{\Theta}\textrm{diag}\left\{ \mathbf{N}_{\mathrm{t}}\right\} \mathbf{\Theta}^{\mathrm{H}}\mathbf{H}_{\mathrm{r}}\!\!+\!\!\kappa_{\mathrm{s}}\left(1\!\!-\!\!\rho^{2}\right)\mathbf{H}_{\mathrm{r}}^{\mathrm{H}}\mathbf{\Theta}\textrm{diag}\left\{ \mathbf{N}_{\mathrm{r}}\right\}\mathbf{\Theta}^{\mathrm{H}}\mathbf{H}_{\mathrm{r}} \nonumber\\
	& \!\!\!\!\!+\!\!\kappa_{\mathrm{d}}\left(1\!\!-\!\!\rho^{2}\right)\!\textrm{diag}\{ \mathbf{H}_{\mathrm{r}}^{\mathrm{H}}\mathbf{\Theta}\textrm{diag}\!\left\{ \mathbf{N}_{\mathrm{t}}\right\} \!\mathbf{\Theta}^{\mathrm{H}}\mathbf{H}_{\mathrm{r}}\} ]^{-1}.
\end{align}
~~Secondly, with given  equalizer $\mathbf{C}$ and  RIS reflecting matrix $\mathbf{\Theta}$, we optimize the procoder $ \mathbf{W} $. However, it is difficult to optimize $ \mathbf{W} $ directly  because the procoder $ \mathbf{W} $ exists in the trace of Eq.\eqref{a4} has different forms in terms of the tiers of the `diag' calculation,  such as $ \text{diag}\{\mathbf{W}\mathbf{W}^\mathrm{H}\} $, $ \text{diag}\{\mathbf{H}_{\mathrm{t}}\text{diag}\{\mathbf{W}\mathbf{W}^\mathrm{H}\}\mathbf{H}_{\mathrm{t}}^\mathrm{H}\} $ and $ \textrm{diag}\left\{ \mathbf{H}_{\mathrm{r}}^{\mathrm{H}}\mathbf{\Theta}\textrm{diag}\left\{ \!\mathbf{H}_{\mathrm{t}} \mathbf{W}\mathbf{W}^{\mathrm{H}} \mathbf{H}_{\mathrm{t}}^{\mathrm{H}}\right\} \!\mathbf{\Theta}^{\mathrm{H}}\mathbf{H}_{\mathrm{r}}\right\}$.
Then we derive two siginificant properties $\text{tr}\{\mathbf{A}\text{diag}\{\mathbf{B}\}\mathbf{C}^\mathrm{H}\}=\text{tr}\{\mathbf{B}\text{diag}\{\mathbf{C}^\mathrm{H}\mathbf{A}\}\}$ and $ \text{tr}\{\mathbf{A}\text{diag}\{\mathbf{B}\text{diag}\{\mathbf{C}\}\mathbf{B}^\mathrm{H}\}\mathbf{A}^\mathrm{H}\}=\text{tr}\{\mathbf{C}\text{diag}\{\mathbf{B}^\mathrm{H}\text{diag}\{\mathbf{A}^\mathrm{H}\mathbf{A}\}\mathbf{B}\}\} $.
Based on these properties, we can optimize the optimization variable $ \mathbf{W} $ in \eqref{a4} by separating it from the diagonal transformation. Then Problem \eqref{a5} can be formulated as
 \begin{align}	\label{a8}
 		\underset{\mathbf{W}}{\min} &~~ \text{MSE}_{w}\\\nonumber
 		\text{s.t.} &\ \text{tr}\{\mathbb{E}[\tilde{\mathbf{x}}\tilde{\mathbf{x}}^\mathrm{H}]\}\leq \tau,
 \end{align}
where  
\begin{align*}
		\!\!\!\text{tr}\{\mathbb{E}[\tilde{\mathbf{x}}\tilde{\mathbf{x}}^\mathrm{H}]\}
		\!\!=\!\!\text{tr}\{ \mathbf{W}\mathbf{W}^\mathrm{H} \!\!+\!\!\kappa_\mathrm{s} \text{diag}\{\mathbf{W}\mathbf{W}^\mathrm{H} \}\!\}\!\!=\!\!(1\!\!+\!\!\kappa_\mathrm{s}) \text{tr}\{\mathbf{W}\mathbf{W}^\mathrm{H}  \} ,
\end{align*}
and $	\text{MSE}_{w}$ is given in \eqref{rr}   at the top of this page.
\begin{figure*}[!t]
	\begin{align}\label{rr}
		 \text{MSE}_{w}&\overset{(a)}{=}\!\text{tr}\{\!\mathbf{W}\mathbf{W}^{\mathrm{H}}\tilde{\mathbf{H}}^{\mathrm{H}}\!\!\mathbf{C}^{\mathrm{H}}\!\mathbf{C}\tilde{\mathbf{H}}\!\!+\!\!\kappa_{\mathrm{s}}\mathbf{W}\mathbf{W}^{\mathrm{H}}\text{diag}\{\tilde{\mathbf{H}}^{\mathrm{H}}\!\mathbf{C}^{\mathrm{H}}\!\mathbf{C}\tilde{\mathbf{H}}\}+\sigma_{\mathrm{n}}^{2}\mathbf{C}\mathbf{C}^{\mathrm{H}}+\kappa_{\mathrm{d}}\mathbf{W}\mathbf{W}^{\mathrm{H}}\tilde{\mathbf{H}}^{\mathrm{H}}\text{diag}\{\mathbf{C}^{\mathrm{H}}\mathbf{C}\}\tilde{\mathbf{H}}-\mathbf{C}\tilde{\mathbf{H}}\mathbf{W}-\mathbf{W}^{\mathrm{H}}\tilde{\mathbf{H}}^{\mathrm{H}}\mathbf{C}^{\mathrm{H}}\}\nonumber\\
		& \!+\!\left(1-\rho^{2}\right)\textrm{tr}\left\{ \mathbf{W}\mathbf{W}^{\mathrm{H}}\mathbf{H}_{\mathrm{t}}^{\mathrm{H}}\textrm{diag}\left\{ \mathbf{\Theta}^{\mathrm{H}}\mathbf{H}_{\mathrm{r}}\mathbf{C}^{\mathrm{H}}\mathbf{C}\mathbf{H}_{\mathrm{r}}^{\mathrm{H}}\mathbf{\Theta}\right\} \mathbf{H}_{\mathrm{t}}\right\} \!+\!\kappa_{\mathrm{s}}\left(1\!-\!\rho^{2}\right)\textrm{tr}\left\{ \mathbf{W}\mathbf{W}^{\mathrm{H}}\textrm{diag}\left\{ \mathbf{H}_{\mathrm{t}}^{\mathrm{H}}\textrm{diag}\left\{ \mathbf{\Theta}^{\mathrm{H}}\mathbf{H}_{\mathrm{r}}\mathbf{C}^{\mathrm{H}}\mathbf{C}\mathbf{H}_{\mathrm{r}}^{\mathrm{H}}\mathbf{\Theta}\right\}\mathbf{H}_{\mathrm{t}}\right\}  \right\}\nonumber \\
		& +\kappa_{\mathrm{d}}\left(1-\rho^{2}\right)\textrm{tr}\left\{ \mathbf{W}\mathbf{W}^{\mathrm{H}}\mathbf{H}_{\mathrm{t}}^{\mathrm{H}}\textrm{diag}\left\{ \mathbf{\Theta}^{\mathrm{H}}\mathbf{H}_{\mathrm{r}}\textrm{diag}\left\{ \mathbf{C}^{\mathrm{H}}\mathbf{C}\right\}\mathbf{H}_{\mathrm{r}}^{\mathrm{H}}\mathbf{\Theta}\right\}  \mathbf{H}_{\mathrm{t}}\right \},
	\end{align}
\vspace{-2.7em}
\rule[1.5ex]{2.05\columnwidth}{0.5pt}
\end{figure*}
Obviously, Problem \eqref{a8} is convex, and its  optimal $\mathbf{W}^{\mathrm{opt}}$ can be obtained by resorting to the Karush-Kuhn-Tucker (KKT) conditions. 
Specifically, the Lagrangian function of Problem \eqref{a8} is given by
\begin{align*}
\mathcal L&=
\text{MSE}_{w}+\lambda(\text{tr}\{\mathbb{E}[\tilde{\mathbf{x}}\tilde{\mathbf{x}}^\mathrm{H}]\}-\tau)\\
&=\text{MSE}_{w}+\lambda(1+\kappa_\mathrm{s})(\text{tr}\{\mathbf{W}\mathbf{W}^\mathrm{H}\}-\frac{\tau}{1+\kappa_\mathrm{s}}),
\end{align*}
where $\lambda$ is the Lagrangian multiplier. The KKT conditions can be formulated as
\begin{subequations}\label{a10}
	\begin{align}
		&\frac{\partial\mathcal L}{\partial \mathbf{W}^*}=0,\label{3}\\
		&\lambda \ge 0,~~~\text{tr}\{\mathbf{W}\mathbf{W}^\mathrm{H}\}\leq \frac{\tau}{1+\kappa_\mathrm{s}}\label{a12},\\
		&\lambda(1+\kappa_\mathrm{s})(\text{tr}\{\mathbf{W}\mathbf{W}^\mathrm{H}\}-\frac{\tau}{1+\kappa_\mathrm{s}})=0\label{a11}.	
	\end{align}
\end{subequations}
Then \eqref{3} can be formulated as
\[
\begin{aligned}\frac{\partial\mathcal{L}}{\partial\mathbf{W}^{*}}\!\overset{(b)}{=} & \tilde{\mathbf{H}}^{\mathrm{H}}\mathbf{C}^{\mathrm{H}}\mathbf{C}\tilde{\mathbf{H}}\mathbf{W}\!+\!\kappa_{\mathrm{s}}\text{diag}\{\tilde{\mathbf{H}}^{\mathrm{H}}\mathbf{C}^{\mathrm{H}}\mathbf{C}\tilde{\mathbf{H}}\}\mathbf{W}\!+\!\kappa_{\mathrm{d}}\tilde{\mathbf{H}}^{\mathrm{H}}\\
	& \text{diag}\{\mathbf{C}^{\mathrm{H}}\mathbf{C}\}\tilde{\mathbf{H}}\mathbf{W}\!+\!\lambda(1\!+\!\kappa_{\mathrm{s}})\mathbf{W}\!-\!\tilde{\mathbf{H}}^{\mathrm{H}}\mathbf{C}^{\mathrm{H}}\!+\!\left(1\!-\!\rho^{2}\right)\\
	& \mathbf{H}_{\mathrm{t}}^{\mathrm{H}}\textrm{diag}\{\mathbf{\Theta}^{\mathrm{H}}\mathbf{H}_{\mathrm{r}}\mathbf{C}^{\mathrm{H}}\mathbf{C}\mathbf{H}_{\mathrm{r}}^{\mathrm{H}}\mathbf{\Theta}\}\mathbf{H}_{\mathrm{t}}\mathbf{W}+\kappa_{\mathrm{s}}\left(1-\rho^{2}\right)\\
	& \!\textrm{diag}\{ \mathbf{H}_{\mathrm{t}}^{\mathrm{H}}\textrm{diag}\{\mathbf{\Theta}^{\mathrm{H}}\mathbf{H}_{\mathrm{r}}\mathbf{C}^{\mathrm{H}}\mathbf{C}\mathbf{H}_{\mathrm{r}}^{\mathrm{H}}\mathbf{\Theta}\}\mathbf{H}_{\mathrm{t}}\} \!\mathbf{W}\!\!+\!\!\kappa_{\mathrm{d}}\left(1\!\!-\!\!\rho^{2}\right)\\
	& \mathbf{H}_{\mathrm{t}}^{\mathrm{H}}\textrm{diag}\{ \mathbf{\Theta}^{\mathrm{H}}\mathbf{H}_{\mathrm{r}}\textrm{diag}\{ \mathbf{C}^{\mathrm{H}}\mathbf{C}\}\mathbf{H}_{\mathrm{r}}^{\mathrm{H}}\mathbf{\Theta}\}  \mathbf{H}_{\mathrm{t}}\mathbf{W},
\end{aligned}
\]
where (b) is from  $\frac{\partial\text{tr}\{ \mathbf{Z}\mathbf{A}\mathbf{Z^H}\mathbf{B}\}}{\partial \mathbf{Z^*}}=\mathbf{B}\mathbf{Z}\mathbf{A} $.
Thus, we can obtain the closed-form solution of the optimal $ \mathbf{W} $ as
\begin{align}\label{a13}
	\mathbf{W}^{\mathrm{opt}}\!\!=
	 [\mathbf{A}\!+\!\lambda(1\!+\!\kappa_{\mathrm{s}})\mathbf{I}]^{-1}\tilde{\mathbf{H}}^{\mathrm{H}}\mathbf{C}^{\mathrm{H}},
\end{align}
where 
\begin{align*}
	\mathbf{A}\!= & \tilde{\mathbf{H}}^{\mathrm{H}}\mathbf{C}^{\mathrm{H}}\mathbf{C}\tilde{\mathbf{H}}\!+\!\kappa_{\mathrm{s}}\text{diag}\{\tilde{\mathbf{H}}^{\mathrm{H}}\mathbf{C}^{\mathrm{H}}\mathbf{C}\tilde{\mathbf{H}}\}\!+\!\kappa_{\mathrm{d}}\tilde{\mathbf{H}}^{\mathrm{H}}\text{diag}\{\mathbf{C}^{\mathrm{H}}\mathbf{C}\}\!\tilde{\mathbf{H}}\\
	& \!+\left(1-\rho^{2}\right)\mathbf{H}_{\mathrm{t}}^{\mathrm{H}}\textrm{diag}\{ \mathbf{\Theta}^{\mathrm{H}}\mathbf{H}_{\mathrm{r}}\mathbf{C}^{\mathrm{H}}\mathbf{C}\mathbf{H}_{\mathrm{r}}^{\mathrm{H}}\mathbf{\Theta}\} \mathbf{H}_{\mathrm{t}}+\kappa_{\mathrm{s}}(1-\rho^{2})\\
	& \textrm{diag}\{ \mathbf{H}_{\mathrm{t}}^{\mathrm{H}}\textrm{diag}\{ \mathbf{\Theta}^{\mathrm{H}}\mathbf{H}_{\mathrm{r}}\mathbf{C}^{\mathrm{H}}\mathbf{C}\mathbf{H}_{\mathrm{r}}^{\mathrm{H}}\mathbf{\Theta}\}\mathbf{H}_{\mathrm{t}}\}  +\kappa_{\mathrm{d}}(1-\rho^{2})\mathbf{H}_{\mathrm{t}}^{\mathrm{H}}\\
	& \textrm{diag}\{ \mathbf{\Theta}^{\mathrm{H}}\mathbf{H}_{\mathrm{r}}\textrm{diag}\{ \mathbf{C}^{\mathrm{H}}\mathbf{C}\}\mathbf{H}_{\mathrm{r}}^{\mathrm{H}}\mathbf{\Theta}\}  \mathbf{H}_{\mathrm{t}}.
\end{align*}

As shown in \eqref{a13}, $\mathbf{W}^{\mathrm{opt}}$ depends on the Lagrangian multiplier $\lambda$. 
It can be found that if both $\lambda=0$ and $\text{tr}\{\mathbf{W}^{\mathrm{opt}}{\mathbf{W}^{\mathrm{opt}}}^\mathrm{H}\}-\frac{\tau}{1+\kappa_\mathrm{s}}\leq 0$ are  satisfied,  $\lambda=0$. Otherwise, $\lambda$ should be the solution to the equation $\text{tr}\{\mathbf{W}^{\mathrm{opt}}{\mathbf{W}^{\mathrm{opt}}}^\mathrm{H}\}=\frac{\tau}{1+\kappa_\mathrm{s}}$. 
In the following part, we will discuss the method of updating $\lambda$ depending on whether $\mathbf{A}$ is full rank or not. 

\subsubsection{Case 1}
If $\mathbf{A}$ is full rank, it can be decomposed by the singular value decomposition  $\mathbf{A}=\mathbf{B}\mathbf{S}\mathbf{B}^\mathrm{H}$, where $\mathbf{B}$ and $\mathbf{S}$ are a unitary matrix and a diagonal matrix, respectively. Then we have
	\begin{align}\label{30}
	\mathbf{W}^{\mathrm{opt}}
	&=(\mathbf{B}\mathbf{S}\mathbf{B}^\mathrm{H}+\lambda(1+\kappa_\mathrm{s})\mathbf{I})^{-1}\tilde{\mathbf{H}}^\mathrm{H}\mathbf{C}^\mathrm{H}\\\nonumber
	&=\mathbf{B}(\mathbf{S}+\lambda(1+\kappa_\mathrm{s})\mathbf{I})^{-1}\mathbf{B}^\mathrm{H}\tilde{\mathbf{H}} ^\mathrm{H}\mathbf{C}^\mathrm{H}.
	\end{align}
The power contraint can be formulated as
\begin{subequations}\label{a31}
	\begin{align}
	&\text{tr}\{\mathbf{W}\mathbf{W}^\mathrm{H}\}\\\nonumber
	&=\!\text{tr}\left\{\!\mathbf{B}(\mathbf{S}\!\!+\!\!\lambda(1+\kappa_\mathrm{s})\mathbf{I})^{\!-1}\mathbf{B}^\mathrm{H}\tilde{\mathbf{H}} ^\mathrm{H} \!\mathbf{C}^\mathrm{H} \mathbf{C}\tilde{\mathbf{H}}\mathbf{B}(\mathbf{S}\!+\!\lambda(1\!\!+\!\!\kappa_\mathrm{s})\mathbf{I})^{\!-1}\mathbf{B}^\mathrm{H} \right\}\\
	&=\text{tr}\left\{(\mathbf{S}+\lambda(1\!\!+\!\!\kappa_\mathrm{s})\mathbf{I})^{-2}\mathbf{B}^\mathrm{H}\tilde{\mathbf{H}} ^\mathrm{H}\mathbf{C}^\mathrm{H}\mathbf{C}\tilde{\mathbf{H}}\mathbf{B}\right\}\\
	&=\text{tr}\left\{(\mathbf{S}+\lambda(1+\kappa_\mathrm{s})\mathbf{I})^{-2}\mathbf{Z}\right\}\\
	&=\sum_{i=1}^{N_\mathrm{t}}\frac{[\mathbf{Z}]_{i,i}}{([\mathbf{S}]_{i,i}+\lambda(1+\kappa_\mathrm{s}))^{2}}\leq \frac{\tau}{1+\kappa_\mathrm{s}},\label{a32}
	\end{align}
\end{subequations}
where $[\mathbf{Z}]_{i,i}$ and $[\mathbf{S}]_{i,i}$ denote the $i$-th diagonal element of $\mathbf{Z}$ and $\mathbf{S}$, respectively. From \eqref{a32}, $\text{tr}\left\{\mathbf{W}\mathbf{W}^\mathrm{H}\right\}$  is monotonically decreasing with $\lambda$. Therefore, if $\text{tr}\left\{\mathbf{W}\mathbf{W}^\mathrm{H}\right\}-\frac{\tau}{1+\kappa_\mathrm{s}}\textgreater 0$, we can use the bisection  search method to obtain the optimal $\lambda$. The upper bound of $\lambda$ can be given by
	\begin{align}\label{a33}
	\lambda  \textless \sqrt{\frac{\sum_{i=1}^{N_\mathrm{t}}[\mathbf{Z}]_{i,i}}{ \frac{\tau}{1+\kappa_\mathrm{s}}}}\bigg/(1+\kappa_\mathrm{s})\triangleq \lambda_{ub}.
	\end{align}
\subsubsection{Case 2}
If $\mathbf{A}$ is not full rank, \eqref{30} cannot be applied. Thus, we verify whether $\lambda=0$ is the optimal solution or not. If $\text{tr}\left\{\mathbf{W}\mathbf{W}^\mathrm{H}\right\}-\frac{\tau}{1+\kappa_\mathrm{s}}\leq 0$, the optimal $\mathbf{W}^{\mathrm{opt}}$ can be rewritten as
	\begin{align}\label{a35}
		\mathbf{W}^{\mathrm{opt}}=\mathbf{A}^{-1}\tilde{\mathbf{H}} ^\mathrm{H}\mathbf{C}^\mathrm{H}.
	\end{align}
Otherwise, the optimal solution of $\lambda$ can be obtained by the bisection search  method.

\subsection{Phase Shift Optimization}
In this section, we optimize the RIS phase shift matrix $\mathbf{\Theta}$ with given equalizer $\mathbf{C}$ and  precoder $\mathbf{W}$. 
Similar to the procoder $\mathbf{W}$, we first separate the RIS phase shift matrix $\mathbf{\Theta}$ from the diagonal transformation.
Then we rewrite Problem \eqref{a5} as
\begin{subequations}\label{a14}
	\begin{align}
	& \underset{\mathbf{\Theta}}{\min} ~~ \text{MSE}_{\theta}\label{uuu}\\
	&\text{s.t.} 
	\ 0\leq\theta_m\leq2\pi,m=1,\cdots,M,
	\end{align}
\end{subequations}
where $	\text{MSE}_{\theta}$ is given in \eqref{a15}   at the top of  this page.

\begin{figure*}[!t]
	\begin{align}\label{a15}
		\textrm{MSE}_{\theta} & =\rho^{2}\textrm{tr}\left(\mathbf{C}\mathbf{H}_{\mathrm{r}}^{\mathrm{H}}\mathbf{\Theta}\mathbf{H}_{\mathrm{t}}\mathbf{W}\mathbf{W}^{\mathrm{H}}\mathbf{H}_{\mathrm{t}}^{\mathrm{H}}\mathbf{\Theta}^{\mathrm{H}}\mathbf{H}_{\mathrm{r}}\mathbf{C}^{\mathrm{H}}\right)+\left(1-\rho^{2}\right)\textrm{tr}\left(\mathbf{C}\mathbf{H}_{\mathrm{r}}^{\mathrm{H}}\mathbf{\Theta}\text{diag}\{\mathbf{H}_{\mathrm{t}}\mathbf{W}\mathbf{W}^{\mathrm{H}}\mathbf{H}_{\mathrm{t}}^{\mathrm{H}}\}\mathbf{\Theta}^{\mathrm{H}}\mathbf{H}_{\mathrm{r}}\mathbf{C}^{\mathrm{H}}\right)+\rho\,\,\textrm{tr}(\mathbf{H}_{\mathrm{t}}\mathbf{W}\mathbf{W}^{\mathrm{H}}\nonumber\\
		& \!\mathbf{H}_{\mathrm{d}}\mathbf{C}^{\mathrm{H}}\mathbf{C}\mathbf{H}_{\mathrm{r}}^{\mathrm{H}}\mathbf{\Theta}\!+\!\mathbf{\Theta}^{\mathrm{H}}\mathbf{H}_{\mathrm{r}}\mathbf{C}^{\mathrm{H}}\mathbf{C}\mathbf{H}_{\mathrm{d}}^{\mathrm{H}}\mathbf{W}\mathbf{W}^{\mathrm{H}}\mathbf{H}_{\mathrm{t}}^{\mathrm{H}})\!\!+\!\!\rho\kappa_{\mathrm{s}}\textrm{tr}(\mathbf{H}_{\mathrm{t}}\text{diag}\{\mathbf{W}\mathbf{W}^{\mathrm{H}}\}\mathbf{H}_{\mathrm{d}}\mathbf{C}^{\mathrm{H}}\mathbf{C}\mathbf{H}_{\mathrm{r}}^{\mathrm{H}}\mathbf{\Theta}\!\!+\!\!\mathbf{\Theta}^{\mathrm{H}}\mathbf{H}_{\mathrm{r}}\mathbf{C}^{\mathrm{H}}\mathbf{C}\mathbf{H}_{\mathrm{d}}^{\mathrm{H}}\text{diag}\{\mathbf{W}\mathbf{W}^{\mathrm{H}}\}\mathbf{H}_{\mathrm{t}}^{\mathrm{H}})\nonumber\\
		& \!+\rho^{2}\kappa_{\mathrm{s}}\,\textrm{tr}\left(\mathbf{C}\mathbf{H}_{\mathrm{r}}^{\mathrm{H}}\mathbf{\Theta}\mathbf{H}_{\mathrm{t}}\text{diag}\{\mathbf{W}\mathbf{W}^{\mathrm{H}}\}\mathbf{H}_{\mathrm{t}}^{\mathrm{H}}\mathbf{\Theta}^{\mathrm{H}}\mathbf{H}_{\mathrm{r}}\mathbf{C}^{\mathrm{H}}\right)+\,\kappa_{\mathrm{s}}\left(1-\rho^{2}\right)\textrm{tr}(\mathbf{C}\mathbf{H}_{\mathrm{r}}^{\mathrm{H}}\mathbf{\Theta}\text{diag}\{\mathbf{H}_{\mathrm{t}}\text{diag}\{\mathbf{W}\mathbf{W}^{\mathrm{H}}\}\mathbf{H}_{\mathrm{t}}^{\mathrm{H}}\}\mathbf{\Theta}^{\mathrm{H}}\mathbf{H}_{\mathrm{r}}\mathbf{C}^{\mathrm{H}})\nonumber\\
		& \!+\rho\,\kappa_{\mathrm{d}}\textrm{tr}\left(\mathbf{H}_{\mathrm{d}}^{\mathrm{H}}\mathbf{W}\mathbf{W}^{\mathrm{H}}\mathbf{H}_{\mathrm{t}}^{\mathrm{H}}\mathbf{\Theta}^{\mathrm{H}}\mathbf{H}_{\mathrm{r}}\text{diag}\{\mathbf{C}^{\mathrm{H}}\mathbf{C}\}+\mathbf{H}_{\mathrm{r}}^{\mathrm{H}}\mathbf{\Theta}\mathbf{H}_{\mathrm{t}}\mathbf{W}\mathbf{W}^{\mathrm{H}}\mathbf{H}_{\mathrm{d}}\text{diag}\{\mathbf{C}^{\mathrm{H}}\mathbf{C}\}\right)+\rho^{2}\kappa_{\mathrm{d}}\textrm{tr}(\mathbf{H}_{\mathrm{r}}^{\mathrm{H}}\mathbf{\Theta}\mathbf{H}_{\mathrm{t}}\mathbf{W}\mathbf{W}^{\mathrm{H}}\mathbf{H}_{\mathrm{t}}^{\mathrm{H}}\mathbf{\Theta}^{\mathrm{H}}\mathbf{H}_{\mathrm{r}}\nonumber\\
		& \!\!\!\text{diag}\{\mathbf{C}^{\mathrm{H}}\mathbf{C}\})\!\!+\!\!\kappa_{\mathrm{d}}\!\left(1\!\!-\!\!\rho^{2}\right)\textrm{tr}\!\!\left(\mathbf{H}_{\mathrm{r}}^{\mathrm{H}}\mathbf{\Theta}\text{diag}\{\mathbf{H}_{\mathrm{t}}\mathbf{W}\mathbf{W}^{\mathrm{H}}\mathbf{H}_{\mathrm{t}}^{\mathrm{H}}\}\mathbf{\Theta}^{\mathrm{H}}\mathbf{H}_{\mathrm{r}}\text{diag}\{\mathbf{C}^{\mathrm{H}}\mathbf{C}\}\right)\!\!-\!\!\rho\textrm{tr}\!\!\left(\mathbf{C}\mathbf{H}_{\mathrm{r}}^{\mathrm{H}}\mathbf{\Theta}\mathbf{H}_{\mathrm{t}}\mathbf{W}\!\!+\!\!\mathbf{W}^{\mathrm{H}}\mathbf{H}_{\mathrm{t}}^{\mathrm{H}}\mathbf{\Theta}^{\mathrm{H}}\mathbf{H}_{\mathrm{r}}\mathbf{C}^{\mathrm{H}}\right)\!\!.
	\end{align}
	\vspace{-3.8em}
\end{figure*}
Denote the set of diagonal elements of $\mathbf{\Theta}$ by  $\pmb{\theta}\triangleq [e^{j\theta_1},e^{j\theta_2},\cdots,e^{j\theta_M}]^\mathrm{T}$ and  the collection of diagonal elements of a general matrix $\mathbf{P}$  by $\mathbf{p}=[[\mathbf{P}]_{1,1},\cdots,[\mathbf{P}]_{M,M}]^\mathrm{T}$, we have
\begin{align}	\label{a16}
\text{tr}(\mathbf{\Theta}\mathbf{P})+\text{tr}(\mathbf{\Theta}^\mathrm{H}\mathbf{P}^\mathrm{H})=\pmb{\theta}^\mathrm{T}\mathbf{p}+\mathbf{p}^\mathrm{H}\pmb{\theta}^*=2\text{Re} \{\pmb{\theta}^\mathrm{H}\mathbf{p}^*\}.
\end{align}
We define $\mathbf{\Omega}\triangleq \mathbf{H}_\mathrm{t}\mathbf{W}\mathbf{W}^\mathrm{H}\mathbf{H}_\mathrm{d}\mathbf{C}^\mathrm{H}\mathbf{C}\mathbf{H}_\mathrm{r}^\mathrm{H}, 
\mathbf{\Psi}\triangleq$$ \mathbf{H}_\mathrm{t}\text{diag}\{\mathbf{W}\mathbf{W}^\mathrm{H}\}\mathbf{H}_\mathrm{d}$\\$\mathbf{C}^\mathrm{H}\mathbf{C}\mathbf{H}_\mathrm{r}^\mathrm{H}$, $\mathbf{T}\triangleq\mathbf{H}_\mathrm{t}\mathbf{W}\mathbf{W}^\mathrm{H}\mathbf{H}_\mathrm{d}\text{diag}\{\mathbf{C}^\mathrm{H}\mathbf{C}\}\mathbf{H}_\mathrm{r}^\mathrm{H}$, and $\mathbf{V}\triangleq\mathbf{H}_\mathrm{t}\mathbf{W}\mathbf{C}\mathbf{H}_\mathrm{r}^\mathrm{H}$.
Denote the set of diagonal elements of $ \mathbf{\Omega}, \mathbf{\Psi},\mathbf{T},\mathbf{V}$ by $\pmb{\omega},\pmb{\psi},\mathbf{t},\mathbf{v}$, respectively, given by $\pmb{\omega}=[[\mathbf{\Omega}]_{1,1},\cdots,[\mathbf{\Omega}]_{M,M}]^\mathrm{T}$,
$ \pmb{\psi}=[[\mathbf{\Psi}]_{1,1},\cdots,[\mathbf{\Psi}]_{M,M}]^\mathrm{T} $, $ \mathbf{t}=[[\mathbf{T}]_{1,1},\cdots,[\mathbf{T}]_{M,M}]^\mathrm{T} $ and $ \mathbf{v}=[[\mathbf{V}]_{1,1},\cdots,[\mathbf{V}]_{M,M}]^\mathrm{T} $.

By using 
$\text{tr}(\mathbf{\Theta}^\mathrm{H}\mathbf{B}\mathbf{\Theta}\mathbf{C})=\pmb{\theta}^\mathrm{H}(\mathbf{B}\odot\mathbf{C}^\mathrm{T})\pmb{\theta}$, where $\odot$ is  Hadamard product operation,
the objective function shown in \eqref{a15} can be reformulated as
 \eqref{a17} at the top of this page,
 \begin{figure*}[!t]
 	\begin{align}
 		\label{a17}
 		\textrm{MSE}_{\theta} & =\rho^{2}\pmb{\theta}^{\mathrm{H}}[\mathbf{H}_{\mathrm{r}}\mathbf{C}^{\mathrm{H}}\mathbf{C}\mathbf{H}_{\mathrm{r}}^{\mathrm{H}}\odot(\mathbf{H}_{\mathrm{t}}\mathbf{W}\mathbf{W}^{\mathrm{H}}\mathbf{H}_{\mathrm{t}}^{\mathrm{H}})^{\mathrm{T}}]\pmb{\theta}+\left(1-\rho^{2}\right)\pmb{\theta}^{\mathrm{H}}[\mathbf{H}_{\mathrm{r}}\mathbf{C}^{\mathrm{H}}\mathbf{C}\mathbf{H}_{\mathrm{r}}^{\mathrm{H}}\odot\left(\textrm{diag}\left\{ \mathbf{H}_{\mathrm{t}}\mathbf{W}\mathbf{W}^{\mathrm{H}}\mathbf{H}_{\mathrm{t}}^{\mathrm{H}}\right\} \right)^{\mathrm{T}}]\pmb{\theta}+\rho^{2}\kappa_{\mathrm{s}}\pmb{\theta}^{\mathrm{H}}\nonumber\\
 		& \![\mathbf{H}_{\mathrm{r}}\mathbf{C}^{\mathrm{H}}\mathbf{C}\mathbf{H}_{\mathrm{r}}^{\mathrm{H}}\odot\!(\mathbf{H}_{\mathrm{t}}\textrm{diag}\left\{ \mathbf{W}\mathbf{W}^{\mathrm{H}}\right\} \mathbf{H}_{\mathrm{t}}^{\mathrm{H}})^{\mathrm{T}}]\pmb{\theta}\!+\!\kappa_{\mathrm{s}}\left(1\!-\!\rho^{2}\right)\pmb{\theta}^{\mathrm{H}}[\mathbf{H}_{\mathrm{r}}\mathbf{C}^{\mathrm{H}}\mathbf{C}\mathbf{H}_{\mathrm{r}}^{\mathrm{H}}\odot(\textrm{diag}\left\{ \mathbf{H}_{\mathrm{t}}\textrm{diag}\left\{ \mathbf{W}\mathbf{W}^{\mathrm{H}}\right\} \mathbf{H}_{\mathrm{t}}^{\mathrm{H}}\right\} )^{\mathrm{T}}]\pmb{\theta}\!+\!\rho^{2}\kappa_{\mathrm{d}}\nonumber\\
 		&\! \pmb{\theta}^{\mathrm{H}}[\mathbf{H}_{\mathrm{r}}\textrm{diag}\left\{ \mathbf{C}^{\mathrm{H}}\mathbf{C}\right\} \mathbf{H}_{\mathrm{r}}^{\mathrm{H}}\odot(\mathbf{H}_{\mathrm{t}}\mathbf{W}\mathbf{W}^{\mathrm{H}}\mathbf{H}_{\mathrm{t}}^{\mathrm{H}})^{\mathrm{T}}]\pmb{\theta}+\kappa_{\mathrm{d}}\left(1-\rho^{2}\right)\pmb{\theta}^{\mathrm{H}}[\mathbf{H}_{\mathrm{r}}\textrm{diag}\left\{ \mathbf{C}^{\mathrm{H}}\mathbf{C}\right\} \mathbf{H}_{\mathrm{r}}^{\mathrm{H}}\odot(\textrm{diag}\left\{ \mathbf{H}_{\mathrm{t}}\mathbf{W}\mathbf{W}^{\mathrm{H}}\mathbf{H}_{\mathrm{t}}^{\mathrm{H}}\right\} )^{\mathrm{T}}]\pmb{\theta}\nonumber\\
 		&\! +2\rho\text{Re}\{\pmb{\theta}^{\mathrm{H}}\pmb{\omega}^{*}\}+2\rho\kappa_{\mathrm{s}}\text{Re}\{\pmb{\theta}^{\mathrm{H}}\pmb{\psi}^{*}\}+2\rho\kappa_{\mathrm{d}}\text{Re}\{\pmb{\theta}^{\mathrm{H}}\mathbf{t}^{*}\}-2\rho\text{Re}\{\pmb{\theta}^{\mathrm{H}}\mathbf{v}^{*}\},
 	\end{align}
\vspace{-2.0em}
\rule[1.5ex]{2.05\columnwidth}{0.5pt}
 \end{figure*}
where
\begin{align*}
	\mathbf{\Xi}= & \rho^{2}[\mathbf{H}_{\mathrm{r}}\mathbf{C}^{\mathrm{H}}\mathbf{C}\mathbf{H}_{\mathrm{r}}^{\mathrm{H}}\!\odot\!(\mathbf{H}_{\mathrm{t}}\mathbf{W}\mathbf{W}^{\mathrm{H}}\mathbf{H}_{\mathrm{t}}^{\mathrm{H}})^{\mathrm{T}}]\!\!+\!\!\left(1\!-\!\rho^{2}\right)[\mathbf{H}_{\mathrm{r}}\mathbf{C}^{\mathrm{H}}\mathbf{C}\\
	& \mathbf{H}_{\mathrm{r}}^{\mathrm{H}}\odot(\textrm{diag}\{ \mathbf{H}_{\mathrm{t}}\mathbf{W}\mathbf{W}^{\mathrm{H}}\mathbf{H}_{\mathrm{t}}^{\mathrm{H}}\} )^{\mathrm{T}}]+\rho^{2}\kappa_{\mathrm{s}}[\mathbf{H}_{\mathrm{r}}\mathbf{C}^{\mathrm{H}}\mathbf{C}\mathbf{H}_{\mathrm{r}}^{\mathrm{H}}\odot\\
	& (\mathbf{H}_{\mathrm{t}}\textrm{diag}\{\mathbf{W}\mathbf{W}^{\mathrm{H}}\}\mathbf{H}_{\mathrm{t}}^{\mathrm{H}})^{\mathrm{T}}]+\kappa_{\mathrm{s}}\left(1-\rho^{2}\right)[\mathbf{H}_{\mathrm{r}}\mathbf{C}^{\mathrm{H}}\mathbf{C}\mathbf{H}_{\mathrm{r}}^{\mathrm{H}}\odot\\
	& (\textrm{diag}\{\mathbf{H}_{\mathrm{t}}\textrm{diag}\{\mathbf{W}\mathbf{W}^{\mathrm{H}}\}\mathbf{H}_{\mathrm{t}}^{\mathrm{H}}\})^{\mathrm{T}}]+\rho^{2}\kappa_{\mathrm{d}}[\mathbf{H}_{\mathrm{r}}\textrm{diag}\{ \mathbf{C}^{\mathrm{H}}\mathbf{C}\} \\
	& \mathbf{H}_{\mathrm{r}}^{\mathrm{H}}\odot(\mathbf{H}_{\mathrm{t}}\mathbf{W}\mathbf{W}^{\mathrm{H}}\mathbf{H}_{\mathrm{t}}^{\mathrm{H}})^{\mathrm{T}}]+\kappa_{\mathrm{d}}\left(1-\rho^{2}\right)[\mathbf{H}_{\mathrm{r}}\textrm{diag}\{ \mathbf{C}^{\mathrm{H}}\mathbf{C}\} \\
	& \mathbf{H}_{\mathrm{r}}^{\mathrm{H}}\odot(\textrm{diag}\{ \mathbf{H}_{\mathrm{t}}\mathbf{W}\mathbf{W}^{\mathrm{H}}\mathbf{H}_{\mathrm{t}}^{\mathrm{H}}\} )^{\mathrm{T}}],\\
	\mathbf{Q}= & \pmb{\omega}^{*}+\kappa_{s}\pmb{\psi}^{*}+\kappa_{d}\mathbf{t}^{*}-\mathbf{v}^{*}.
\end{align*}

Define $\phi_{m}=e^{j\theta_m},\forall m$, and $\pmb{\theta}=[\phi_{1},\cdots,\phi_{M}]^\mathrm{T}$.
Problem \eqref{a14} can be rewritten as
\begin{subequations}\label{a19}
		\vspace{-0.5em}
	\begin{align}
	&	\underset{\pmb{\theta}}{\min}  \ f(\pmb{\theta})=\pmb{\theta}^\mathrm{H}\mathbf{\Xi}\pmb{\theta}+2\rho\text{Re} \{\pmb{\theta}^\mathrm{H}\mathbf{Q}\}\\
	&	\text{s.t.} 
		\ |\phi_{m}|=1, m=1,\cdots,M.\label{33}
	\end{align}
\end{subequations}
Due to the unit modulus constraints in \eqref{33},  Problem \eqref{a19} is a nonconvex problem. In the following, we provide a Majorization-Minimization (MM) algorithm to solve this problem. 
Specifically,  we construct a function $g(\pmb{\theta}|\pmb{\theta}^t)$ as an upper bound of the original objective function $f(\pmb{\theta})$. Denote the 
solution of $f(\pmb{\theta})$  and the objective fuction value of Problem \eqref{a19} at the $t$-th iteration  by $\pmb{\theta}^t$ and $f(\pmb{\theta}^t)$, respectively. 
Then, the remaining problem is to determine the upper bound objective function $g(\pmb{\theta}|\pmb{\theta}^t)$. 
It can be observed from  Problem \eqref{a19} that  $\mathbf{\Xi}$  is a Herimitian matrix. Utilizing the $ Claim  $~1 of \cite{7}, we have 
\begin{align}\label{a22}
	\!\!\!	\!\!\!\!\!\pmb{\theta}^\mathrm{H}\mathbf{\Xi}\pmb{\theta} \leq \pmb{\theta}^\mathrm{H}\mathbf{\Lambda}\pmb{\theta}\!\!+\!\!2\text{Re} \{\pmb{\theta}^\mathrm{H}(\mathbf{\Xi}\!\!-\!\!\mathbf{\Lambda})\pmb{\theta}^t\}\!\!+\!\!(\pmb{\theta}^t)^\mathrm{H}\!(\mathbf{\Lambda\!\!-\!\!\mathbf{\Xi}})\pmb{\theta}^t\!\!\triangleq\! y(\pmb{\theta}|\pmb{\theta}^{t}),
\end{align}
where $\mathbf{\Lambda}$ should satisfy $\mathbf{\Lambda} \succeq \mathbf{\Xi}$. Let $\mathbf{\Lambda}=\lambda_{max}(\mathbf{\Xi})\mathbf{I}_M$, where $ \lambda_{max}(\mathbf{\Xi}) $ is the largest eigenvalue of matrix $ \mathbf{\Xi} $. The objective function $g(\pmb{\theta}|\pmb{\theta}^t)$ can be formulated as
$
	g(\pmb{\theta}|\pmb{\theta}^t)= y(\pmb{\theta}|\pmb{\theta}^{t})+2\rho\text{Re} \{\pmb{\theta}^\mathrm{H}\mathbf{Q}\}.
$
The consequence of $\pmb{\theta}^\mathrm{H}\mathbf{\Lambda}\pmb{\theta}$ is a constant  due to $\pmb{\theta}^\mathrm{H}\pmb{\theta}=\mathbf{M}$. In addition, $(\pmb{\theta}^t)^\mathrm{H}(\mathbf{\Lambda-\mathbf{\Xi}})\pmb{\theta}^t $ is also a constant since vector $ \pmb{\theta}^{t} $ is known at the  $t$-th iteration.  Problem \eqref{a19} at the $ (t+1) $-th iteration is given by
\begin{subequations}\label{a24}
\begin{align}
&\underset{\pmb{\theta}}{\max}  \ 2\text{Re} \{\pmb{\theta}^\mathrm{H}\mathbf{u}^t\}\\
&\text{s.t.} 
\ |\phi_{m}|=1, m=1,\cdots,M,
\end{align}
\end{subequations}
where $\mathbf{u}^t=-(\mathbf{\Xi}-\lambda_{max}(\mathbf{\Xi})\mathbf{I}_M)\pmb{\theta}^{t}-\rho\mathbf{Q}$. The closed-form solution  of Problem \eqref{a24} can be derived as
\begin{align}\label{a25}
(\pmb{\theta}^{t+1})^*=e^{j\text{arg}(\mathbf{u}^t)}.
\end{align}
Based on the above subsections, the overall AO algorithm to
solve Problem \eqref{a5} is summarized in Algorithm 1.


%

\begin{algorithm}[htbp]
\caption{:Algorithm to solve Problem \eqref{a5}}
	\label{b2}
	\begin{algorithmic}[1]
		\STATE Initialization: Randomly initialize  $\pmb{\theta}$ and $\mathbf{W}$. Normalize $\mathbf{W}^{1}$ to meet the power constraint and set $\pmb{\Theta}^{1}=\text{diag}\{\pmb{\theta}^1\}$.  Set the iteration number $t=1$ and the convergence accuracy $\epsilon \rightarrow 0$.
		\REPEAT 
		\STATE Update $\mathbf{C}^{t}$ with fixed $\mathbf{W}^{t}  $ and $ \mathbf{\Theta}^{t} $ according to \eqref{a7}.
		\STATE Update $\mathbf{W}^{t+1}$ with fixed $ \mathbf{\Theta}^{t} $ and $\mathbf{C}^{t}$ according to \eqref{a13}.
		\STATE Update $ \pmb{\theta}^{t+1} $ with fixed $\mathbf{W}^{t+1}$ and $\mathbf{C}^{t}$  by \eqref{a25} and recover $ \mathbf{\Theta}^{t+1} $ from $ \pmb{\theta}^{t+1} $, set $t \leftarrow t+1$.
		\UNTIL the difference of MSE in two iterations less than  $\epsilon$.
	\end{algorithmic}
\end{algorithm}

\section{Numerical Results}
In this section, we evaluate the performance of the proposed algorithm for an  RIS-aided single-user MIMO system with HIs impact. The normalized mean squared error (NMSE) is defined as $\frac{\text{MSE}}{\mathbb{E}\{\Vert \mathbf{s} \Vert_{2}^{2}\}}=\frac{\text{MSE}}{d}$. We set the locations of the BS and the RIS as (0 m, 0 m) and (10 m, 0 m), respectively. The large-scale path loss is modeled as $PL=-30-10\alpha \text{log}_{10}(d)$,
where $\alpha$ and  $d$ are the path-loss exponent and the  distance of the transmission link, respectively. 
We set the path-loss exponent of line-of-sight (LoS) channel and non-LoS channel as 2 and 3.75, respectively.
The small-scale fading is assumed to be Rician fading, i.e.,
	$
	\mathbf{H}=\sqrt{\frac{\beta}{\beta+1}}\mathbf{H}^{\mathrm{LoS}}+\sqrt{\frac{1}{\beta+1}}\mathbf{H}^{\mathrm{NLoS}},
	$
	where $ \beta $ is the Rician factor, $\mathbf{H}^{\mathrm{LoS}}  $ and $ \mathbf{H}^{\mathrm{NLoS}} $ represent the deterministic line of sight (LoS) and  the non-LoS (NLoS) components, respectively.
	The $\mathbf{H}^{\mathrm{LoS}}  $ is given by $ \mathbf{H}^{\mathrm{LoS}}=\mathbf{a}_{r}(\iota_r,\gamma_r)\mathbf{a}_{t}(\iota_t,\gamma_t)^\mathrm{H}  $, where $ 	\mathbf{a}(\iota,\gamma) $ is the steering vector of uniform planar array (UPA) with 
	$\iota_r $  (resp. $\gamma_r$), and $ \iota_t  $ (resp. $ \gamma_t $)  representing the azimuth (resp. elevation)
	angles of departure and arrival for the LoS component, respectively.
The other simulation parameters are set as follows:  Rician factor of $\beta=10$; thermal noise power density of -104 dBm/Hz, system bandwidth of $B=1$ MHz, the number of transmit antennas of $N_\mathrm{t}=8$; the number of receive antennas of $N_\mathrm{r}=4$; the number of data steams of $d=4$;  the parameter of the phase noise of $\rho=20$, and the error tolerance of $e=10^{-5}$. In order to show the performance of the proposed system model more clearly, the following five  schemes are compared: 
1) RIS with HIs: the proposed design;
2) RIS without HIs: conventional  design in an RIS-aided system without  HIs;
3) No-RIS with HIs: conventional transceiver design considering HIs impact in
a single-user MIMO system without an RIS; 
4) Randphase-RIS with HIs: the phase shifts of the reflecting elements are randomly set and we only optimize the beamforming matrices at the BS and user;
5) Naive design: in this scheme, we consider an RIS-aided system with HIs. However, we adopt the existing methods that ignore the HIs. Then, we plug the obtained solutions into the actual RIS-aided system with HIs. This scheme is used to demonstrate the advantages of considering HIs in the system design.

\begin{figure}[t]
	\centering
	\centerline{\includegraphics[width=3in]{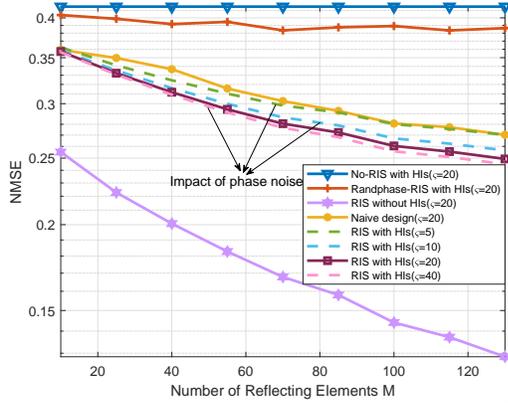}}
	\vspace{-1.2em}
\caption{Downlink achievable NMSE versus the number of RIS elements $M$.} 
	\vspace{-1.77em}
\end{figure}

Fig.~1 shows the NMSE of different schemes versus the number of RIS reflecting elements when $\kappa_\mathrm{s}=0.1$ and $\kappa_\mathrm{d}=0.1$. 
It can be seen that the NMSE value decreases with the increase of  the number of reflecting elements $M$. It is because that increasing $M$ can improve the channel conditions between the BS and the user. 
The randphase RIS sheme shows an overall downward trend with the increase of $M$.
Besides, the NMSE is  stable with the increase of $M$ for \!the \!no-RIS \!scheme, and its value is much higher than our proposed \!scheme. This demonstrates the advantage of deploying an RIS in a communication system.
Furthermore, the proposed scheme  outperforms the naive design. The reason is that the naive design does not take the HIs into consideration that exists in real commnunication systems. 
It reveals that ignoring HIs will result in system performance degradation, thereby indicating the significance of transceiver design for RIS-aided systems by considering HIs.
In addition, the dashed lines show the impact of RIS phase shift noise on the downlink  NMSE. It is obvious that a higher $\varsigma$  leads to reduced NMSE.  It is because that
	$ \varepsilon $ decreases as $\varsigma$ increases.
	When $ \varsigma \rightarrow \infty $,  then $ \varepsilon \rightarrow 0$, $\rho \rightarrow 1 $ which means the ideal  RIS hardware.

\begin{figure}[t]
	\centering
	\centerline{\includegraphics[width=3.0in]{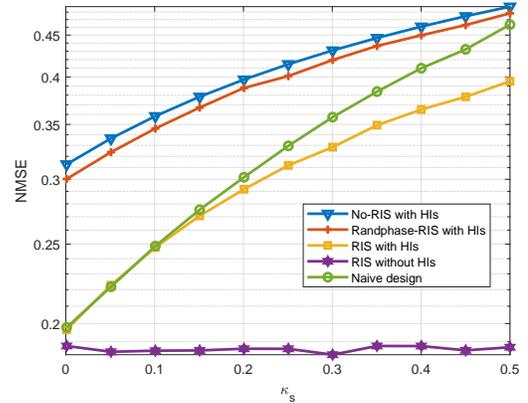}}
	\vspace{-1.2em}
	\caption{Achievable NMSE  versus the transmit distortion noise coefficient $\kappa_\mathrm{s}$.} 
	\label{figsystem}
	\vspace{-1.7em}
\end{figure}
Fig.~2 compares the NMSE of various shemes verus $\kappa_\mathrm{s}$ when $M=40$.
We can find that the NMSE performance deteriorates as $\kappa_\mathrm{s}$ increases by taking into account the HIs. 
This is because the distortion noise power caused by hardware  increases when $\kappa_\mathrm{s}$ is large,  which leads to large NMSE value.
The proposed RIS-aided sheme  outperforms randphase RIS sheme and no-RIS sheme under the same $\kappa_\mathrm{s}$ when considering HIs.
In addition, compared with the perfect hardware conditions, a large $\kappa_\mathrm{s}$ brings a large NMSE performance loss.
This illustrates that it is important to consider HIs in RIS-aided communication systems.
Furthermore, the NMSE is nearly stable with the increase of $\kappa_\mathrm{s}$ for the RIS-aided scheme without incorporating the HIs.

\section{Conclusions}
In this paper, we  investigated the transceiver design for an RIS-aided single-user MIMO communication system with imperfect
HIs.
We aimed to minimize the MSE
 by  jointly optimizing  the precoding matrix at the transmitter, decoding matrix at the receiver  and the phase shifts matrices at the RIS.
To tackle this non-convex problem, the  algorithm based on the Lagrangian dual method and MM algorithm was proposed.
Simulation results showed that the efficiency of our proposed algorithm and the
significance of considering HIs in the transceiver design for RIS-aided system.

\bibliographystyle{IEEEtran}
\bibliography{ref}

\end{document}